**Ultrathin and ultrasmooth gold films on monolayer MoS$_2$**

*Dmitry I. Yakubovsky, Yury V. Stebunov, Roman V. Kirtaev, Georgy A. Ermolaev, Mikhail S. Mironov, Sergey M. Novikov, Aleksey V. Arsenin, and Valentyn S. Volkov\**

D. I. Yakubovsky, Dr. Y. V. Stebunov, R. V. Kirtaev, G. A. Ermolaev, M. S. Mironov, Dr. S. M. Novikov, Dr. A. V. Arsenin, Prof. V. S. Volkov
Center for Photonics and 2D Materials
Moscow Institute of Physics and Technology
Dolgoprudny 141700, Russia

Dr. Y. V. Stebunov, Dr. A. V. Arsenin, Prof. V. S. Volkov
GrapheneTek
Skolkovo Innovation Center
Moscow 143026, Russia

Prof. V. S. Volkov
SDU Nano Optics
Mads Clausen Institute
University of Southern Denmark
Campusvej 55, DK-5230, Odense, Denmark
E-mail: vsv@mci.sdu.dk

Sub-10 nm continuous metal films are promising candidates for flexible and transparent nanophotonics and optoelectronics applications. In this Letter, we demonstrate that monolayer MoS$_2$ is a perspective adhesion layer for the deposition of continuous conductive gold films with a thickness of only 3-4 nm. Optical properties of continuous ultrathin gold films deposited on two-dimensional MoS$_2$ grown by chemical vapor deposition are investigated by spectroscopic ellipsometry over a wide wavelength range (300-3300 nm). Results show that optical losses in ultrathin films increase with decreasing thickness due to the fine-grained structure and the presence of a small number of voids, however, they exhibit metallic properties down to a thickness of 3-4 nm. The atomic-scale MoS$_2$ interfaces can be transferred to any substrate and thus open up new opportunities for the creation of metasurfaces and a new type of van der Waals heterostructures with atomically thin metal layers.

Gold is the most widely used metal for plasmonic and metamaterial applications due to its relatively low optical losses in the visible and NIR ranges, high electrical conductivity and good chemical stability.[1-3] Nowadays thin gold films are an important component of highly efficient photonic and plasmonic devices.[4] Ultrathin gold films (thickness <10 nm) have recently attracted a great deal of interest both for the development of flexible transparent electrodes for optoelectronic devices[5-8] including thin-film solar cells, displays and touchscreens, photodetector and light emitting diodes, and for studying quantum-size effects in metal films.[9] Certainly, high-quality ultrathin films are also the key element of plasmonic waveguides[10,11] and hyperbolic metamaterials.[12,13]

Growth of continuous and ultrathin gold films on different substrates, such as glass, silicon oxide, silicon nitride, graphene etc. is notoriously difficult due to the poor wetting of gold to these substrates.[14,15] The growth kinetics of metal films is generally determined by the adsorption and diffusion behavior of metal adatoms on the substrate. A small ratio of the adsorption energy of metal adatoms on the substrate to the bulk cohesive energy of the metal and low diffusion barrier for an adatom favor the three-dimensional island growth behavior also known as the Volmer-Weber growth mode.[16] Within the framework of this growth model, the formation of a metal film is associated with the following stages: nucleation of islands, island growth, island impingement and coalescence, percolation, and channel-filling to finally form a continuous thin film. To reduce the percolation threshold of ultrathin gold films, adhesion or seed layers of Ti, Cr, Ni, Pt or Ge are commonly used. However, these adhesion layers significantly affect the optical and electrical properties of ultrathin metal nanostructures.[17-22] Recently, the organosilane-based adhesion layers (mercaptosilanes and aminosilanes) were used for the deposition of sub-10-nm-thick continuous Au films on silicon and glass surfaces.[23-27] However, organosilanes are not compatible with non-oxidized silicon surfaces and poorly compatible with standard lift-off procedures, that imposes severe limitations to their use as adhesion layers.[21] Adhesion layers based on organosilanes are also inefficient for the deposition of atomically thin metal films[14] and does not move us closer to the deposition of two-dimensional layers from bulk plasmonic metals. Actually, the latter seems now as impossible as the deposition of atomically thin carbon films had been considered before 2004.[28]

In the present paper, we propose the use of MoS$_2$ monolayer as an entirely new type of "universal" (i.e., it can be transferred to any arbitrary substrate) adhesion layer for ultrathin (<10 nm) high-quality continuous gold films. To test the feasibility of this idea, we deposited ultrathin gold films of different thicknesses onto monolayer MoS$_2$, grown on silicon dioxide substrates (**Figure 1**a), and studied their structural and optical properties.

An electron beam evaporator Nano Master NEE-4000 was used to deposit Au films on top of atmospheric pressure CVD (APCVD)-grown full area coverage MoS$_2$ monolayers on silicon wafers with a 285 nm SiO$_2$ coating (from 2D semiconductors Inc.). The deposition was performed at room temperature using gold pellets with a purity of 99.999% (Kurt J. Lesker). The base pressure in the vacuum chamber before the evaporation process was as low as ≈ 5×10$^{-7}$ Torr, and it increased to ≈ 2×10$^{-6}$ Torr during evaporation. The high deposition rate of about 5 Å/s and a resulted mass film thickness were controlled during deposition by a quartz-crystal sensor mounted in the vacuum chamber. The thin Au films thickness was also independently determined by step height atomic force microscopy (AFM) measurements (NT-MDT Ntegra Aura) and the root-mean-square (RMS) surface roughness value was determined for each film (Figure 1b). The degree of the SiO$_2$/Si substrate coverage by MoS$_2$ was estimated through optical microscopy (Figure 1c). Monolayer MoS$_2$ film covers more than 97% of the substrate area (1 cm × 1 cm). The quality of MoS$_2$ monolayers was assessed with a Horiba LabRAM HR Evolution confocal Raman microscope. The measurements (Figure 1d) were conducted with a visible laser light (λ = 632.8 nm) at low incident power levels, typically less than ~1.8 mW, measured with a 1800 lines/mm diffraction grating and a 100×/N.A.=0.90 microscope objective. The statistics were collected from (at least) 25 points from different parts of the sample. The position of Raman modes ($E^1_{2g}$ and $A_{1g}$) of $MoS_2$ located near 385 cm$^{-1}$ and 404 cm$^{-1}$, corresponds to a single-layer $MoS_2$[29,30] and indicates the high overall quality of the sample.

Spectroscopic ellipsometry (SE) has been applied to study the optical properties of monolayer $MoS_2$ (Figure 1e) and ultrathin Au films deposited on monolayer $MoS_2$ samples and pure $SiO_2$/Si substrates. The dielectric function spectra were evaluated from data measured using a variable-angle spectroscopic ellipsometer (WVASE®, J. A. Woollam Co.) operating in the wavelength range of 300 – 3300 nm. Parameters of Ψ and Δ were measured at angles of incidence of 60°, 65° and 70°. To investigate the nanomorphology of the films we used the scanning electron microscopy (SEM, JEOL JSM-7001F). Finally, sheet resistances of all deposited Au films were measured by the four-point probe system (Jandel RM3000).

Scanning electron micrographs in **Figure 2** demonstrate the surface morphology of Au films of different thicknesses on $MoS_2$ and $SiO_2$ substrates (obtained at the same deposition process). The difference in film morphologies can be characterized by the metal filling fraction. As shown in Figure 2, Au films on $SiO_2$ primarily consist of isolated metal nanoislands merging into a percolating film with multiple voids at the film thickness of 9 nm. While, in comparison, Au films on $MoS_2$ exhibit structure of closely linked clusters at a thickness of 2.1 nm and almost continuous structure with tiny voids at 4.1 nm. Such differences in structure are a result of different kinetic processes of Au film growth. Thus, the interaction of metal atoms with a substrate is determined by the ratio of the adsorption energy $E_a$ of the metal atom on the substrate to the metal cohesive energy $E_c$ and the diffusion barrier $E_d$.[31] In the case of Au on $MoS_2$, the first-principles calculations based on periodic density functional theory predict a fairly low value of $E_a/E_c = 0.22$,[32] which means that the Au film growth mechanism obeys Volmer-Weber mode with the formation of islands at the initial stage of the film growth. However, $E_d = 0.28$ eV is rather large,[32] that limits to some extent the Au adatoms diffusion on the MoS$_2$ surface and, therefore, leads to the formation of a large number of nucleation centers followed by the constitution of the fine grain structure of the film with the percolation threshold below 3 nm (according to SEM data in Figure 2).

As well as SEM results, a significant difference in structural morphology of Au films grown on MoS$_2$ and SiO$_2$ substrates is also observed in the analysis of AFM scans presented in **Figure 3**. Here, AFM imaging demonstrates that gold films of 4.1 nm on MoS$_2$ are more densely packed and smoother than those formed on SiO$_2$. Thus, RMS surface roughness value of 0.22 nm is determined for Au on MoS$_2$ (with a three-fold decrease in the average roughness compared to SiO$_2$-based films). In addition, measurements carried out with the films of different thicknesses (2.1 – 9.0 nm) showed that RMS roughness doesn't exceed 0.35 nm for films on MoS$_2$ and 0.81 nm on SiO$_2$. Recently, similar features on the surface morphology were reported for 2 nm-thick gold clusters (with RMS roughness value of 0.36 nm) on isolated monolayer MoS$_2$ triangles on SiO$_2$/Si substrate.[33] Concerning our study, we have obtained an RMS roughness of 0.34 nm for the same thickness of the continuous Au film (Figure 1b).

In **Figure 4**, we show effective dielectric functions of ultrathin Au films. They were obtained using an approach described in our previous paper.[34,35] Additionally, in order to accurately determine $\varepsilon'$ and $\varepsilon''$ of gold films by ellipsometry, we obtained dielectric functions of monolayer MoS$_2$ films (Figure 1e) and included them into the ellipsometric model. The Tauc-Lorentz (TL) oscillator model[36] was used to describe interband absorption in monolayer MoS$_2$.[37] Four TL oscillators were used to represent $\varepsilon'$ and $\varepsilon''$ values through the Kramers-Kronig integration. The best fitting results are plotted in Figure 1e and found to be consistent with previously reported data.[37] Furthermore, due to the fact that monolayer MoS$_2$ do not form covalent bonds with Au we must take into account the presents of a van der Waals (vdW) interlayer gap (Figure 1a), which varies from 2.36 Å to 2.74 Å for the Au-MoS$_2$ interface.[32,38]

The results on effective permittivities of gold films on $MoS_2$, shown in Figure 4, demonstrate that an increase in the film thickness from 2.1 to 9.0 nm is accompanied by a significant change in both real $\varepsilon'$ and imaginary $\varepsilon''$ dielectric parts. In details, if for the most thin film with of 2.1 nm the resulting effective real part $\varepsilon'$ is positive, then already for films of 3.0 nm and thicker it becomes negative, as is typical for percolated and continuous gold films. Transition of $\varepsilon'$ from a positive value (insulator) to negative (metal) determines the film percolation threshold,[39] that is confirmed by the electrical measurements, showing the non-conducting behavior of 2.1 nm-thick film and conductive from 3.0 nm, and analysis of the morphology carried out by SEM in Figure 2. The thickness dependence of effective dielectric functions correlates well with a metal filling fraction of films, i.e. the fewer number of voids in the metal, the greater the magnitude of the real part and dispersion of $\varepsilon'$ and $\varepsilon''$ gradually becomes closer to the Drude free-electron model of bulk metal as the thickness increases. Comparison of the effective dielectric functions of gold films on $MoS_2$ and on $SiO_2$ in Figure 4 shows lower magnitudes of $\varepsilon'$ for the most thicker film (9 nm) on $SiO_2$ (dotted curve) than for Au 3.0 nm film on $MoS_2$. Such a low metallic optical response of the film on $SiO_2$ substrate is due to high content of voids (18%) in its structure, it is also confirmed by extremely high sheet resistance of this film ~600 Ω/sq. Meanwhile, gold ultrathin films of 3.0 – 9.0 nm deposited on $MoS_2$ exhibit lower electrical resistance from ~300 to 11 Ω/sq decreasing with thickness increase.

In summary, we have proposed to utilize continuous large-area monolayer $MoS_2$ as an effective adhesion layer for growth of ultrathin metal films by a standard deposition method. By extensive optical, electrical and structural characterization we have demonstrated that ultrathin gold films on $MoS_2$ exhibit continuous smooth morphology and Drude plasmonic response at thicknesses down to 3 – 4 nm. The results obtained can advance the use of high quality ultrathin Au films for fabrication of vdW heterostructures[40] and as transparent electrodes for optoelectronic devices. Further research is required to study the influence of the MoS$_2$ monolayers crystallinity (e.g., by use of mechanically exfoliated MoS$_2$ flakes) on growth, structural, optical and electrical properties of thin metal films. Furthermore, it is of great interest to explore the impact of other transition metal dichalcogenides-based adhesion monolayers on the growth and characteristics of ultrathin continuous metal films.

**Acknowledgements**
This research was funded by Russian Science Foundation, grant number 18-19-00684. We thank Dr. Andrey Vyshnevyy for fruitful discussions. We thank the Shared Facilities Center of the Moscow Institute of Physics and Technology (grant no. RFMEFI59417X0014) for the use of their equipment. D.I.Y. and Y.V.S. contributed equally to this work and should be considered co-first authors.

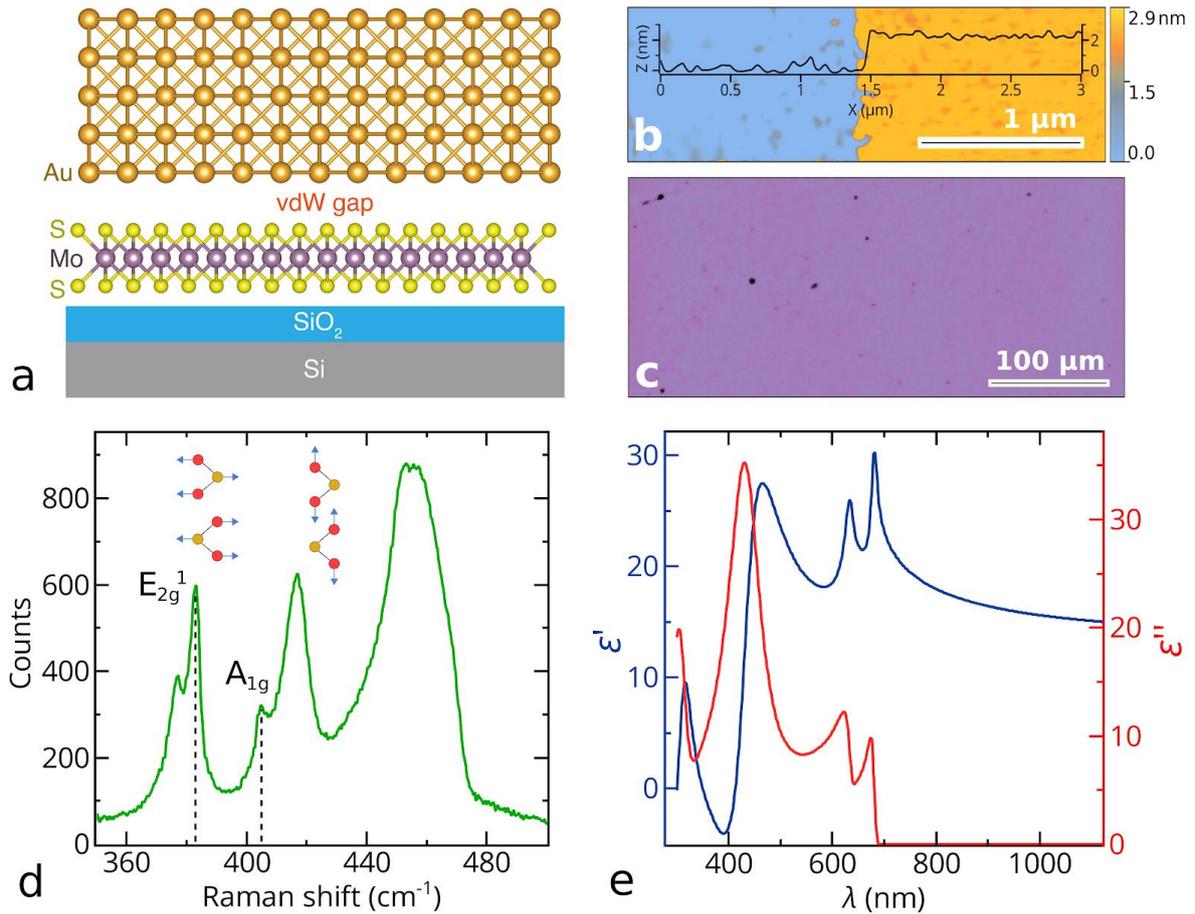

**Figure 1.** (a) A schematic illustration of the Au/MoS$_2$/SiO$_2$/Si structure. (b) The AFM image of the scratch on the 2.1-nm-thick gold film deposited on single-layer MoS$_2$ and the height profile of the scratched gold film. RMS value of Au film roughness is 0.34 nm. (c) An optical microscopy image of a full area coverage MoS$_2$ monolayer on a SiO$_2$/Si substrate. (d) The Raman spectrum of single-layer MoS$_2$ directly grown onto the SiO$_2$/Si substrate. (e) The measured real $\varepsilon'$ (blue line) and imaginary $\varepsilon''$ (red line) parts of the dielectric functions of single-layer MoS$_2$ on the SiO$_2$/Si substrate.

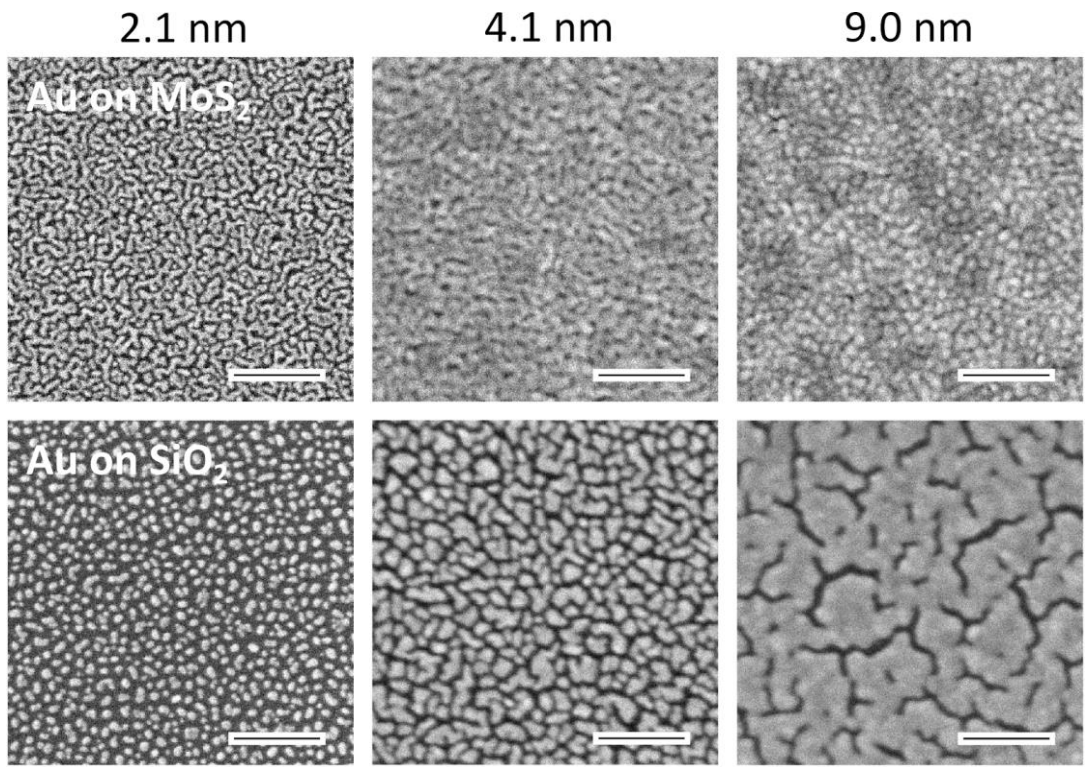

**Figure 2.** SEM images (scale bar = 100 nm) of the surface morphologies of gold film on monolayer $MoS_2$ (top row) and $SiO_2$/Si substrate (bottom row) at different deposition thickness. All images are 400 nm across.

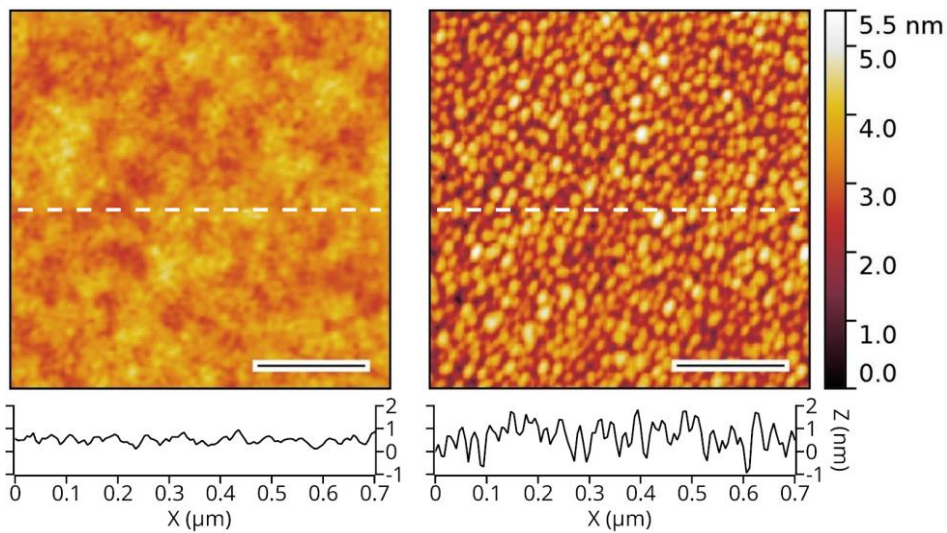

**Figure 3.** AFM images of 4.1 nm gold film on monolayer $MoS_2$ (left) and $SiO_2$/Si substrate (right). Scale bar = 200 nm. The derived RMS values of roughness are 0.22 nm and 0.61 nm for film on monolayer $MoS_2$ and $SiO_2$/Si substrate, respectively.

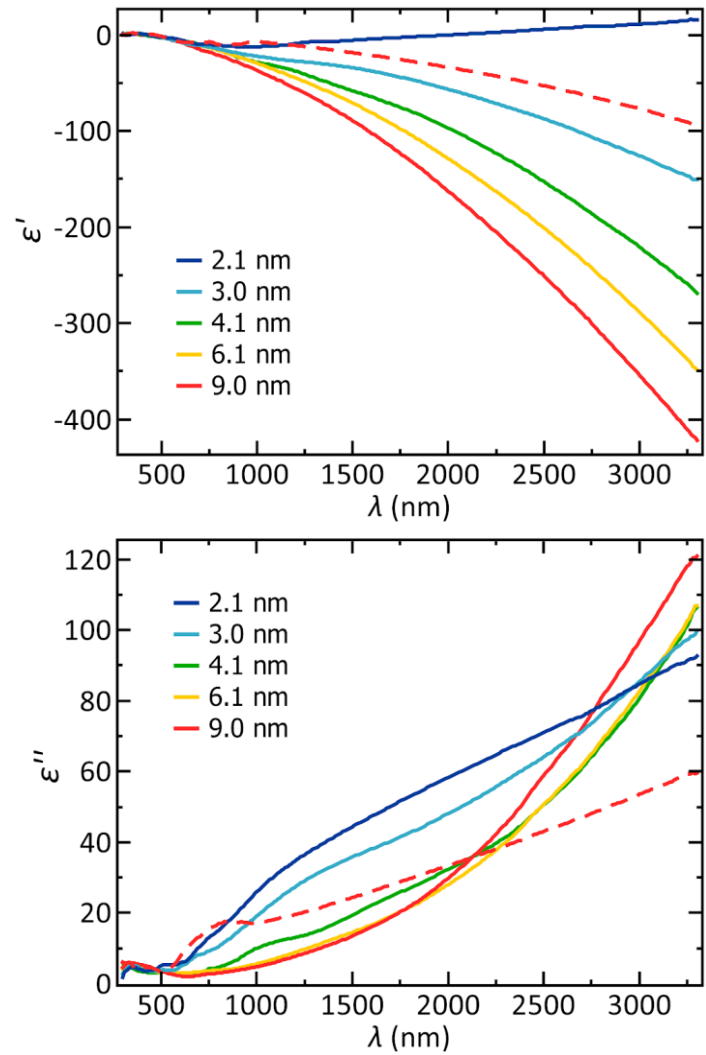

**Figure 4.** The measured real $\varepsilon'$ and imaginary $\varepsilon''$ parts of the dielectric functions of Au films on monolayer $MoS_2$ for several selected thicknesses. The dotted lines show the dielectric function of Au film on $SiO_2$ that was deposited in one cycle together with a 9 nm thick Au film on a $MoS_2$.